\documentclass{appolb}
\usepackage{graphicx}
\def\Pom{{\bf I\!P}}

\begin{document}
\title{Peripheral, ultrarelativistic production of particles \\
in heavy ion collisions
\thanks{Presented at the XX Cracow EPIPHANY Conference on the Physics at the LHC}%
}
\author
{Antoni Szczurek
\address{Institute of Nuclear Physics PAN, PL-31-342 Cracow, Poland and\\
University of Rzesz\'ow, PL-35-959 Rzesz\'ow, Poland}
}
\maketitle

\begin{abstract}
The cross sections for the production of two-pions 
in ultraperipheral ultrarelativistic heavy ion collisions, calculated 
in the impact parameter Equivalent Photon Approximation (EPA), are
presented. 
Differential distributions in impact parameter, dipion invariant mass,
single pion and dipion rapidity, pion transverse momentum and 
pion pseudorapidity are shown.
The $\gamma \gamma \to \pi^+ \pi^-$ subprocess constitutes a background
to the $A A \to A \rho^0 (\to \pi^+ \pi^-) A$ process, 
initiated by emission of a photon by one of colliding nuclei.
Only a part of the dipion invariant mass distribution 
associated with the $\gamma\gamma$-collisions can be visible as 
the cross section for the $A A \to A \rho^0 A$ reaction is very large.

Differential distributions for two $\rho^0$ meson
production in exclusive ultraperipheral, ultrarelativistic collisions
via a double scattering mechanism are presented. 
The cross section for $\gamma A \to \rho^0 A$ is
parametrized based on a calculation from the literature. 
Smearing of $\rho^0$ masses is taken into account. 
The results of calculations are compared to experimental data
obtained at RHIC and to the contribution of the two-photon mechanism.

The double scattering mechanism populates larger $\rho^0 \rho^0$ 
invariant masses and larger rapidity distances between 
the two $\rho^0$ mesons compared to the $\gamma \gamma$ fusion. 
It gives a significant contribution to 
the $A A \to A A \pi^+ \pi^- \pi^+ \pi^-$ reaction.
Some observables related to charged pions are presented.
The results of our calculation are compared with the 
STAR collaboration results for four charged pion production.
While the shape in invariant mass of the four-pion system 
is very similar to the measured one, the predicted cross section 
constitutes only 20 \% of the measured one.
\end{abstract}

\PACS{25.75.Dw,13.25.-k}

%----------------------------  
\section{Introduction}
%----------------------------

Ultrarelativistic ultraperipheral production of mesons and elementary
particles is a special class of nuclear reactions \cite{nucl_reac}.
The ultrarelativistic heavy ions provide large fluxes of quasi-real
photons.
The photon emitted by one nucleus can collide with the other
nucleus or another photon emitted by the second nucleus leading 
to different interesting final states. 
Both total photon-photon cross sections as well as
cross section for particular simple final states are interesting.
The nuclear cross section is often calculated 
in the Equivalent Photon Approximation in momentum space 
\cite{Serbo, KS_muon}.
Alternatively one uses impact parameter space approach \cite{BF90,KS_muon}.
The impact parameter space is very convenient to exclude cases when 
both heavy ions collide with each other, i.e., when they do not survive 
the high-energy collision. 
In the momentum space calculation (EPA or full calculation) the effects 
of nucleus-nucleus collisions and their associated break-up are neglected.
This effect is very small for light particle production such as 
$e^+ e^-$ but increases when the mass of the produced system becomes large.

In our past publications we have shown there the inclusion 
of realistic charge form factors, being Fourier transforms 
of realistic charge distributions, is essential for precise estimate 
of the nuclear cross sections. 

In this presentation, we present results obtained within 
the impact parameter EPA for the exclusive production 
of $\pi^+ \pi^-$ and $\pi^0 \pi^0$. 
For $\rho^0 \rho^0$ production till recently only the photon-photon 
mechanism was discussed in the literature \cite{KN1998,GM2003,GMS2006,KS_rho}.
In Ref. \cite{KS_rho} we have made a first realistic estimate
of the corresponding cross section.

In this mini review we discuss also exclusive production of
two $\rho^0$ mesons. 
The cross section for single$\rho^0$ meson production was predicted 
in the literature to be large \cite{KN1999,FSZ2003,GM2005}. 
Measurements performed at RHIC confirmed
the size of the cross section \cite{STAR2008_rho0},
but were not able to distinguish between different models that predicted
different behaviour on (pseudo)rapidity.
The large cross section for single $\rho^0$ production means
that the cross section for double scattering process is also
large. The best example of a similar type of reaction is the production
of $c \bar c c \bar c$ final state in proton-proton collisions 
which was measured recently by the LHCb
collaboration \cite{LHCb_ccbarccbar}. In our studies \cite{ccbarccbar}  
we predicted and explained the main trends of the data as 
a double-parton scattering effect.
There, the cross section for the $c \bar c c \bar c$ final state is
of the same order of magnitude as the cross section for single 
$c \bar c$ pair production.
The situation for exclusive $\rho^0$ production is similar.
Due to good control of absorption effect, 
the impact parameter formulation seems in the latter case 
to be the best approach.

In Ref.\cite{KS2014} we have studied differential single particle 
distributions for the $\rho^0$ mesons, as well as correlations between 
the $\rho$ mesons, also for the photon-photon component
where also a comparison to the results for photon-photon process 
was done, 
in order to understand how to identify the double photoproduction process. 
In this analysis we have take into account the decay of $\rho^0$ mesons 
into charged pions, in order to take into account some
experimental cuts of existing experiments.
We shall discuss here how to identify the double-scattering 
mechanism at the LHC.

%-----------------------
\section{Formalism}
%-----------------------

In this section we sketch the formalism necessary to understand
exclusive production of $\pi^+ \pi^-$, $\pi^0 \pi^0$,
$\rho^0 \rho^0$ and $\pi^+ \pi^- \pi^+ \pi^-$.

%-----------------------------------------------------------------
\subsection{$\gamma \gamma \to \pi^+ \pi^-$ and 
        $\gamma \gamma \to \pi^0 \pi^0$ processes}
%-----------------------------------------------------------------

\begin{figure}[!h]             
\begin{center}
\includegraphics[width=6cm]{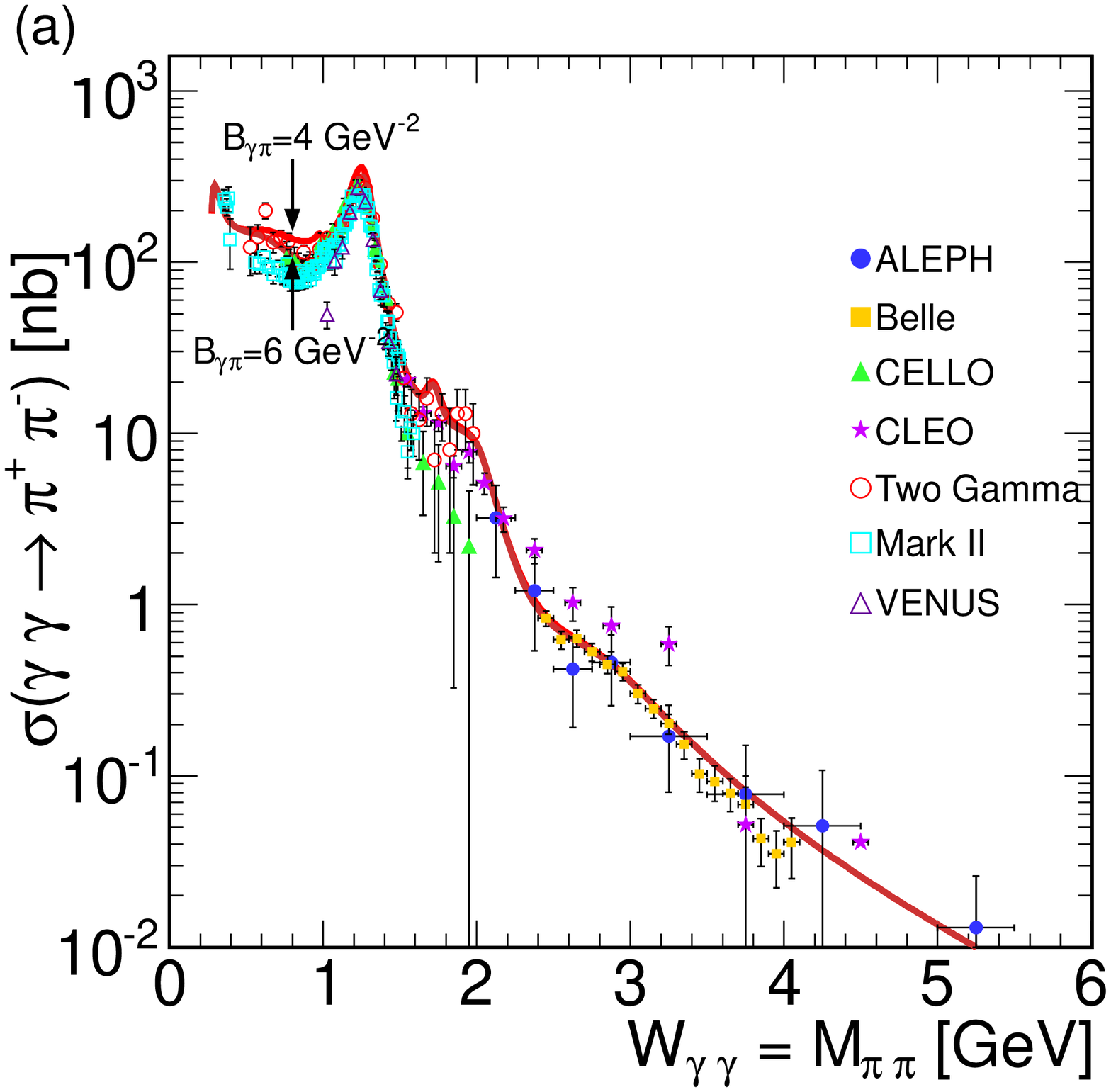}
\includegraphics[width=6cm]{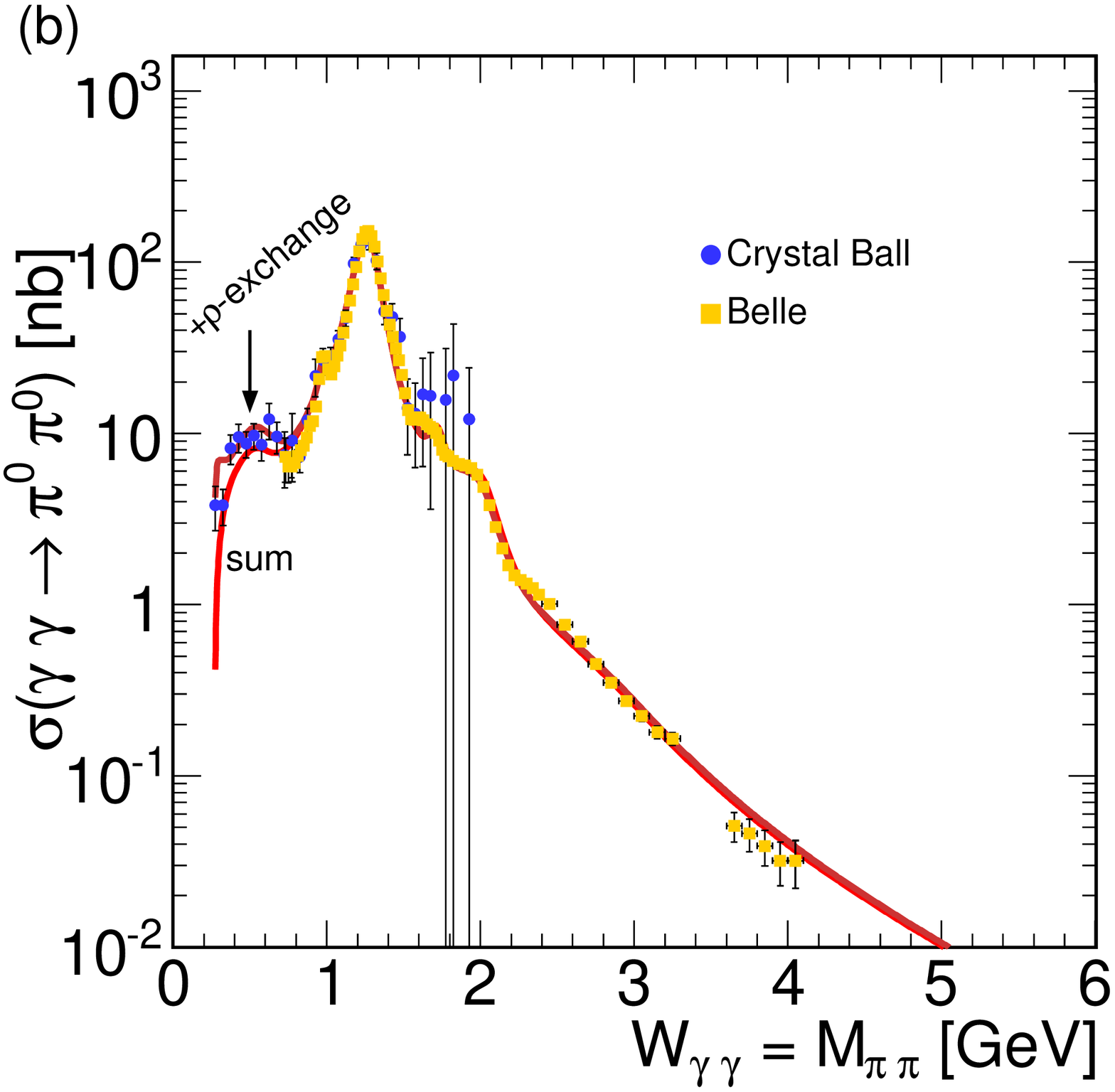}
\end{center}
\caption{\label{fig:gamgam_to_pipi}
\small (Color online) The total cross section for
$\gamma\gamma\to\pi^+\pi^-$ (left panel, $|\cos\theta|<$ 0.6) 
and $\gamma\gamma\to\pi^0\pi^0$ (right panel, $|\cos\theta|<$ 0.8).
}
\end{figure}

In Ref.\cite{KS2013} we have made a careful analysis of reaction
mechanisms necessary to understand the situation for
``elementary'' processes $\gamma \gamma \to \pi^+ \pi^-$
and $\gamma \gamma \to \pi^0 \pi^0$. Here we shall not
repeat details of that study. Instead in Fig.\ref{fig:gamgam_to_pipi} 
we show how we described corresponding world data for these processes.

%--------------------------------------------------------------------
%---------------------------------------------------------------

As discussed in the next section the elementary cross sections 
for $\gamma \gamma \to \pi \pi$ are building blocks for calculating 
corresponding cross sections for nuclear collisions.
In the following we shall concentrate exclusively on the nuclear 
collisions.

%-----------------------------------------------------------------
\subsection{Equivalent photon approximation for 
$AA\to AA\pi\pi$}
%-----------------------------------------------------------------

The total nuclear cross section for these processes can be expressed 
by folding the $\gamma\gamma\to\pi\pi$ subprocess cross sections  
with equivalent photon fluxes as:
\begin{eqnarray}
\sigma(AA\to AA\pi\pi;s_{AA} ) & = & 
\int \hat{\sigma}(\gamma\gamma\to\pi\pi;W_{\gamma\gamma}) S^2_{abs}({\bf b}) \nonumber \\ 
& \times & N(\omega_1,{\bf b}_1) N(\omega_2,{\bf b}_2) 
d^2{\bf b}_1 d^2{\bf b}_2 d\omega_1 d\omega_2 \; .
\label{eq.first_cross_section}
\end{eqnarray}
At intermediate energies the Glauber approach is a reasonable approach
to calculate the absorption factor $S_{abs}({\bf b})$.
Here we approximate the absorption factor as:
\begin{equation}
S^2_{abs}({\bf b}) = \theta({\bf b}-2R_A) = 
                     \theta(|{\bf b}_1-{\bf b}_2|-2R_A) \; .
\end{equation}
This form excludes the geometrical configurations when the colliding
nuclei overlap which at high energies leads to their breakup. 

In order to simplify our calculation we use the transformations:
\begin{equation}
\omega_{1/2} = \frac{W_{\gamma\gamma}}{2} e^{\pm Y_{\pi\pi}}
\mbox{, } 
d\omega_1 d\omega_2 = \frac{W_{\gamma\gamma}}{2} dW_{\gamma\gamma} dY_{\pi\pi} \; .
\end{equation}
The formula (\ref{eq.first_cross_section}) can be written now as:
\begin{eqnarray}
\sigma(AA\to AA\pi\pi;s_{AA} ) & = & 
\int \hat{\sigma}(\gamma\gamma\to\pi\pi;W_{\gamma\gamma}) S^2_{abs}({\bf b}) \nonumber \\ 
& \times & N(\omega_1,{\bf b}_1) N(\omega_2,{\bf b}_2) 
\frac{W_{\gamma\gamma}}{2} d^2{\bf b}_1 d^2{\bf b}_2 d W_{\gamma\gamma} d Y_{\pi\pi} \; .
\label{eq.second_cross_section}
\end{eqnarray}

In the above approach one can easily calculate only the total nuclear 
cross section, distributions in rapidity of the pair of pions,
and the invariant mass of the dipions (see e.g. \cite{KS_rho,KS_muon}),
experimental constraints can not be easily imposed.

If one wants to calculate kinematical distributions of 
each of the individual particles (transverse momentum, rapidity, 
pseudorapidity),
or impose corresponding experimental cuts, 
a more complicated calculation is required. 
Then instead of the one-dimensional function 
$\sigma(\gamma \gamma \to \pi \pi;W_i)$, a two-dimensional functions
$\frac{d \sigma(\gamma \gamma \to \pi \pi)}{dz} \left(W_i, z_j\right)$
have to be calculated.
Then an extra integration in $z=\cos\theta$ is required in 
Eqs.~(\ref{eq.first_cross_section}) or (\ref{eq.second_cross_section}), 
which makes the calculation time-consuming. 

Four-momenta of pions in the $\pi \pi$ center
of mass frame can be calculated as:
\begin{eqnarray}
E_{\pi} = \sqrt{\hat{s}}/2. \; , \nonumber \\
p_{\pi} = \sqrt{\frac{\hat{s}}{4} - m_{\pi}^2}  \; , \nonumber \\
p_{t,\pi} = \sqrt{1-z^2} p_{\pi}  \; , \nonumber \\
p_l = z p_{\pi} \; .
\label{pipi_cm_kinematics}
\end{eqnarray}
The rapidity of the pion can be calculated as:
\begin{equation}
y_i = Y_{\pi \pi} + y_{i/\pi \pi} \; ,
\label{rapidities_of_pions}
\end{equation}
where $y_{i/\pi \pi}=y_{i/\pi \pi}(W,z)$ is the rapidity
of one of the pions in the recoil $\pi\pi$ system.
The transverse momenta of pions in both frames of reference are the same.
Other kinematical variables are calculated by adding relativistically
velocities \cite{Hagedorn_book}
\begin{equation}
\vec{v}_i = \vec{V}_{\pi \pi} \oplus \vec{v}_{i/\pi\pi}  \;  ,
\label{adding_velocities}
\end{equation}
where 
\begin{equation}
\overrightarrow{V_{\pi \pi}} = \frac{\overrightarrow{P_{\pi \pi}}}{E_{\pi \pi}}
\label{velocity_of_recoiled_system}
\end{equation}
and from the energy-momentum conservation:
\begin{eqnarray}
E_{\pi \pi}   &=& \omega_1 + \omega_2   \; , \nonumber \\
P_{\pi \pi}^z &=& \omega_1 - \omega_2   \; ,
\end{eqnarray}
the energies of photons can be expressed as:
\begin{eqnarray}
\omega_1 &=& \frac{W_{\gamma \gamma}}{2} \exp ( Y) \; , \nonumber \\
\omega_2 &=& \frac{W_{\gamma \gamma}}{2} \exp (-Y) \; .
\end{eqnarray}

Now the pion velocities can be converted to four-momenta of pions
in the nucleus-nucleus center of mass frame.
Then angles or pseudorapidities of pions can be calculated
and experimental cuts can be imposed.

%----------------------------------------------------------------------
\subsection{Single scattering production of $\rho^0$}
%----------------------------------------------------------------------

\begin{figure}[!h]
\begin{center}
\includegraphics[width=6.0cm]{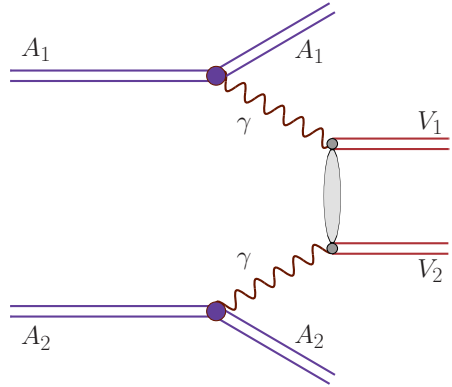}
\end{center}
   \caption{
\small (Color online) The photon-photon mechanism of two vector meson
production in ultrarelativistic ultraperipheral collisions.}
\label{fig:photon_photon}
\end{figure}

Most of the analyses in the literature concentrated on 
production of pairs of mesons in photon-photon processes 
(see Fig. \ref{fig:photon_photon}).
Our group has studied both exclusive $\rho^0 \rho^0$ productions 
\cite{KS_rho} and recently exclusive production of $J/\psi J/\psi$ pairs.

%-----------------------------------------------------------------------------
%------------------------------------------------------------------------------

In the case of double $\rho^0$ production there are two mechanisms.
At larger photon-photon energies the pomeron/reggeon exchange mechanism
is the dominant one, while close to $\rho^0 \rho^0$ threshold a large 
enhancement was observed.
In Ref. \cite{KS_rho} this enhancement of the cross section was
parametrized. In the present paper we shall concentrate rather on 
larger dimeson invariant masses.
 
The elementary cross section for $\gamma \gamma \to \rho^0 \rho^0$
has been measured in the past for not too large energies.
The measured cross section shows a characteristic bump at about
$M_{\rho \rho} \sim$ 1.5 GeV. 
%In Ref. \cite{KS_rho} we parametrized the measured cross 
%section and used the resulting elementary cross section as the input 
%for nuclear calculations.

%In Fig. \ref{fig:possible_mechanisms} we show 
%There are many potential mechanisms contributing to the bump. 
%Since the calculation of the corresponding cross sections is not always easy, 
%we leave theoretical studies
%of the underlying dynamics for future studies. 
At somewhat larger photon-photon energies (larger $\rho^0 \rho^0$ invariant
masses) another mechanism, which can be relatively reasonably
calculated, plays the dominant role. This is a soft (small angle) 
virtual vector meson rescattering.
The corresponding matrix element for small $\rho^0$ meson
transverse momenta can be parametrized in the VDM-Regge approach 
\cite{KS_rho}. 
At large transverse momenta of the $\rho^0$ meson, two-gluon exchange
should become important.
%(see a discussion of two-gluon exchange for 
%$J/\Psi J/\Psi$ production in Ref. \cite{BCKSS2013}). 
In the present analysis we shall discuss only the soft scattering mechanisms.
The hard mechanism may be important at the LHC.
%It is interesting if the two mechanisms can be identified by imposing
%kinematical cuts, or by looking at special observables.

The nuclear cross section for the photon-photon mechanism is calculated
in the impact parameter space as:
\begin{eqnarray}
 \sigma\left( AA \to AA \rho^0\rho^0 \right) &=& 
\int \hat{\sigma} \left( \gamma\gamma \to \rho^0\rho^0 ; W_{\gamma\gamma}\right) 
S^2_{abs}\left( \bf{b} \right) N(\omega_1, {\bf b}_1) N(\omega_2, {\bf b}_2) \nonumber \\
& \times & d^2 {\bf b}_1 d^2 {\bf b}_2 d\omega_1 d\omega_2  \; . 
\end{eqnarray}
Above $W_{\gamma \gamma}$ is the energy in the $\gamma \gamma$ system
and the factor related to absorption is taken as:
\begin{equation}
 S^2_{abs}\left( \bf{b} \right) = \theta ({\bf b}-2R_A) = 
\theta (|{\bf b}_1-{\bf b}_2|-2R_A)  \; .
\end{equation}
This can be written equivalently as:
\begin{eqnarray}
 \sigma\left( AA \to AA \rho^0\rho^0 \right) &=& 
\int \hat{\sigma} \left( \gamma\gamma \to \rho^0\rho^0 ; W_{\gamma\gamma}\right) 
S^2_{abs}\left( \bf{b} \right) N(\omega_1, {\bf b}_1) N(\omega_2, {\bf b}_2) \nonumber \\
& \times & \frac{W_{\gamma\gamma}}{2} d^2 {\bf b}_1 d^2 {\bf b}_2 d
W_{\gamma\gamma} d Y_{\rho^0\rho^0} \; .
\end{eqnarray}
Four-momenta of $\rho^0$ mesons in the $\rho^0\rho^0$ center of mass frame
read:
\begin{equation}
 E_{\rho^0} = \frac{\sqrt{\hat{s}}}{2}  \; ,
\end{equation}
\begin{equation}
 p_{\rho^0} = \sqrt{\frac{\hat{s}}{4} - m_{\rho^0}^2} \; ,
\end{equation}
\begin{equation}
 p_{t,\rho^0} = \sqrt{1-z^2} p_{\rho^0} \; ,
\end{equation}
\begin{equation}
 p_{l,\rho^0} = z p_{\rho^0} \; .
\end{equation}
Above $\hat{s} = W^2_{\gamma \gamma}$ and 
$z = \cos \theta^*$ is defined in the $\rho^0 \rho^0$ center of mass frame. 
In calculations of the photon-photon processes 
the masses of $\rho^0$ mesons are set at their resonance values.

The rapidity of each of the $\rho^0$ mesons ($i$ = 1, 2) is calculated as:
\begin{equation}
 y_{i} = Y_{\rho^0\rho^0} + y_{i/\rho^0\rho^0}(W_{\gamma \gamma},z) \; ,
\end{equation}
where $z$ can be calculated using $\rho^0$ transverse momentum.
$Y_{\rho^0 \rho^0}$ is rapidity of the $\rho^0 \rho^0$ system.

Other kinematical variables are calculated by adding relativistically 
velocities:
\begin{equation}
 \overrightarrow{v}_i = \overrightarrow{v}_{\rho^0\rho^0} \oplus
 \overrightarrow{v}_{i/\rho^0\rho^0}   \; ,
\end{equation}
\begin{equation}
 \overrightarrow{v}_{\rho^0\rho^0} =
 \frac{\overrightarrow{P}_{\rho^0\rho^0}}{E_{\rho^0\rho^0}} \; ,
\end{equation}
where $\vec{v}_{\rho^0 \rho^0}$ is velocity of the $\rho^0 \rho^0$
system in the overall nucleus-nucleus center of mass and 
$\vec{v}_{i/\rho^0 \rho^0}$ is velocity of one of the $\rho^0$ mesons
in the $\rho^0 \rho^0$ system. $\vec{P}_{\rho^0 \rho^0}$ and 
$E_{\rho^0 \rho^0}$ are momentum and energy of the $\rho^0 \rho^0$
system.

The energies of photons is expressed in terms of our integration
variables
\begin{equation}
 \omega_{1/2} = \frac{W_{\gamma\gamma}}{2} \exp (\pm Y_{\rho^0\rho^0})
\end{equation}
from the energy-momentum conservation:
\begin{eqnarray}
 E_{\rho^0\rho^0} &=& \omega_1 + \omega_2 \; , \nonumber \\
 P^z_{\rho^0\rho^0} &=& \omega_1 - \omega_2 \; .
\end{eqnarray}
%
%$Y_{\rho^0 \rho^0}$ is rapidity of the $\rho^0 \rho^0$ system.
The total elementary cross section is calculated as:
\begin{equation}
 \hat{\sigma}(\gamma\gamma\to\rho^0\rho^0) =
 \int^{t_{max}(\hat{s})}_{t_{min}(\hat{s})}
 \frac{d\hat{\sigma}}{d\hat{t}} d\hat{t} \; ,
\end{equation}
where
\begin{equation}
 \frac{d\hat{\sigma}(\gamma\gamma\to\rho^0\rho^0)}{d\hat{t}} = \frac{1}{16\pi\hat{s}^2} |\mathcal{M}_{\gamma\gamma\to\rho^0\rho^0}|^2 \; .
\end{equation}
The matrix element is calculated in a VDM-Regge approach \cite{KS_rho} as
\begin{eqnarray}
{\cal M}_{\gamma \gamma \to \rho^0 \rho^0} &=&
C_{\gamma \to \rho^0} C_{\gamma \to \rho^0} {\hat{s}}
\left( \eta_{\Pom}(\hat{s},\hat{t}) C_{\Pom} 
\left( \frac{\hat{s}}{s_0} \right) ^{\alpha_{\Pom}(t)-1}
+ \eta_{R}(\hat{s},\hat{t}) C_{R} 
\left( \frac{\hat{s}}{s_0} \right)^{\alpha_{R}(t)-1} \right) 
\nonumber \\
&\times& F(\hat{t}, q_1^2 \approx 0) F(\hat{t}, q_2^2 \approx 0) \; .
\label{VDM_Regge_amplitude}
\end{eqnarray}
This is consistent with the existing world experimental data 
on total $\gamma \gamma \to \rho^0 \rho^0$ cross section \cite{KS_rho}.
The photon-to-$\rho^0$ transformation $ C_{\gamma \to \rho^0}$ factor
is calculated in the Vector Dominance Model (VDM). The parameters responsible
for energy dependence are taken from the Donnachie-Landshoff parametrization
of the total proton-proton and pion-proton cross sections \cite{DL92}
assuming Regge factorization. The slope parameter is taken  
as $B$ = 4 GeV$^{-2}$.
The form factors $F(\hat{t}, q^2)$ is described in detail 
in Ref.~\cite{KS_rho}. When calculating kinematical variables, a fixed
resonance position $m_{\rho} = m_R$ is taken for the 
$\gamma \gamma \to \rho^0 \rho^0$.

The differential distributions can be obtained by replacing total
elementary cross section by
\begin{equation}
 \hat{\sigma}(\gamma\gamma\to\rho^0\rho^0) = \int 
\frac{d\hat{\sigma}(\gamma\gamma\to\rho^0\rho^0)}{d {p_t}} d {p_t} \; ,
\end{equation}
where
\begin{equation}
  \frac{d\hat{\sigma}}{d {p_t}}  =  \frac{d\hat{\sigma}}{d {p_t}^2} \frac{d {p_t}^2}{d {p_t}} 
				 =  \frac{d\hat{\sigma}}{d {p_t}^2} 2 p_t 
				 =  \frac{d\hat{\sigma}}{d\hat{t}} 
                       | \partial \hat{t} / \partial p_t^2 | 2 p_t \; .
\label{variable_transformation}
\end{equation}
The first $\rho^0$ is emitted in the forward and the second $\rho^0$
in the backward direction.
% in the $\gamma \gamma \to \rho^0 \rho^0$ center-of-mass system.
The following three-dimensional grids are prepared separately
for the low-energy bump and VDM-Regge components:
\begin{equation}
\frac{d \sigma_{AA \to AA \rho^0 \rho^0}}{dy_1 dy_2 d p_t} \; .
\label{maps}
\end{equation}
The grids are used then to calculate distributions of pions from 
the decays of $\rho^0$ mesons produced in the photon-photon fusion.

%-----------------------------------------------------
\subsection{Single $\rho^0$ production}
%-----------------------------------------------------

The cross section for single vector meson production, differential
in impact factor and vector-meson rapidity, reads:
\begin{equation}
\frac{d \sigma}{d^2 b dy} = 
\omega_1 \frac{d {\tilde N}}{d^2 b d \omega_1} \sigma_{\gamma A_2 \to V
  A_2}(W_{\gamma A_2}) +
\omega_2 \frac{d {\tilde N}}{d^2 b d \omega_2} \sigma_{\gamma A_1 \to V
  A_1}(W_{\gamma A_1}) \; ,
\label{single_vector_meson}
\end{equation}
where $\omega_1 = m_{\rho^0}/2 \exp(+y)$ and 
$\omega_2 = m_{\rho^0}/2 \exp(-y)$.
The flux factor of equivalent photons, ${\tilde N}$, is in principle 
a function of AA impact parameter $b$ and not of 
photon-nucleus impact parameter as is usually done in the literature. 
The effective impact factor can be formally written as the convolution
of real photon flux in one of the nuclei and effective 
strength for interaction of the photon with the second nucleus
\begin{equation}
\frac{d {\tilde N}}{d^2 b d \omega} = \int \frac{d N}{d^2 b_1 d \omega}
\frac{S(b_2)}{\pi R_A^2} d^2 b_1 \approx 
\frac{dN}{d^2b d \omega}  \; ,
\end{equation}
where $\vec{b}_1 = \vec{b} + \vec{b}_2$ and $S(b_2) = \theta(R_A - b_2)$.
It is assumed that the collision occurs when the photon hits
the nucleus. For the photon flux in the second nucleus
one needs to replace 1 $\to$ 2 and 2 $\to$ 1.
 
In general case one can write:
\begin{equation}
\sigma_{\gamma A \to V A}(W) = 
\frac{d \sigma_{\gamma A \to V A}(W,t=0)}{dt} 
\int_{-\infty}^{t_{max}} dt |F_A(t)|^2 \; .
\label{gammaA_VA} 
\end{equation}
Above $W$ is energy in the $\gamma A$ system.
The second factor includes the $t$-dependence for the
$\gamma A \to V A$ subprocess which is due to coherent $q \bar q$ dipole
rescattering off a ``target'' nucleus. This is dictated by 
the nuclear strong form factor.
We approximate the nuclear strong form factor 
by the nuclear charge form factor.
The $t_{max}$ is calculated as 
$t_{max} = - (m_{\rho^0}^2/(2\omega_{lab}))^2$.
$F_A(t)$ is calculated as Fourier transform of the Woods-Saxon 
charge distribution with parameters
specified in Ref. \cite{KS2014}.
The first term in Eq. (\ref{gammaA_VA}) is usually weakly dependent
on the $\gamma A$ energy. For the $\rho^0$ meson it is almost
a constant \cite{KN1999}:
\begin{equation}
\frac{d\sigma(\gamma+A \to \rho^0 A; W,t=0)}{dt} 
\approx \mbox{const} \; .
\label{gammaA_rho0A}
\end{equation}
The constant is taken to be (see \cite{KN1999})
420 mb/GeV$^2$ for RHIC and 450 mb/GeV$^2$ for LHC.
These are cross sections for $W_{\gamma p}$ energies relevant 
for midrapidities at $\sqrt{s_{NN}} =$ 200 GeV and 5.5 TeV,
respectively.

The second term in Eq.(\ref{gammaA_VA}) depends on $t_{max}$ which in turn 
depends rather on running $\rho^0$ meson mass than on resonance position.

The cross section for the $\gamma A \to V A$ reaction could be also
calculated e.g., in the QCD dipole picture in the so-called 
mixed representation (see e.g., \cite{GM2006,Lappi2013}). 
For a more complicated momentum space formulation of the vector meson
production on nuclei see \cite{CSS2012}.

At high energy the imaginary part of the amplitude for the 
$\gamma A \to V A$ process can be expressed as \cite{KNNZ,NNZ}:
\begin{equation}
\Im \left( A_{\gamma A \to V A}(W) \right) =
\Sigma_{\lambda \bar \lambda} \int dz d^2 \rho
\;
\Psi_{\lambda \bar \lambda}^V(z,\rho) 
\;
\sigma_{dip-A}(\rho,W)
\;
\Psi_{\lambda \bar \lambda}^{\gamma}(z,\rho) \; .
\label{general_amplitude_gammaA_VA}
\end{equation}
In the equation above, $\lambda$ and $\bar \lambda$ are 
quark and antiquark helicities. Helicity conservation at high energy 
rescattering of the dipole in the nucleus is assumed.
The variable $\rho$ is the transverse size of the quark-antiquark dipole, 
and $z$/$(1-z)$ denote the longitudinal momentum fractions carried by
quark and antiquark, respectively.
Using explicit formulae for photon and vector meson wave functions, 
the generic formula (\ref{general_amplitude_gammaA_VA}) can be written 
in a simple way.
The dipole-nucleus cross section can be expressed in the
Glauber-Gribov approach in terms of the nuclear thickness
$T_A(b_{\gamma})$, as seen be the $q \bar q$ dipole in its way 
through the nucleus,
and the dipole-proton $\sigma_{dip-p}(\rho)$ cross section as:
\begin{equation}
\sigma_{dip-A}(\rho,W) = 2 \int d^2 b_{\gamma}
\left\{
1 - \exp \left( -\frac{1}{2} T_A(b_{\gamma}) \sigma_{dip-p}(\rho,W) \right)
\right\} \; .
\label{dipole_nucleus_cross_section}
\end{equation}
The formula above allows for an easy way to include
rather complex multiple scattering of the quark-antiquark dipole 
in the nucleus.
Several parametrizations of the dipole-nucleon cross section
were proposed in the literature. 
%Here we shall consider a few simple parametrizations of 
%the dipole-nucleon cross section in order 
%to quantify the related theoretical uncertainties.
Most of them were obtained through fitting HERA deep-inelastic
scattering data. The saturation inspired parametrizations are 
the most popular ones.

\begin{figure}[!h]
\begin{center}
\includegraphics[width=6cm]{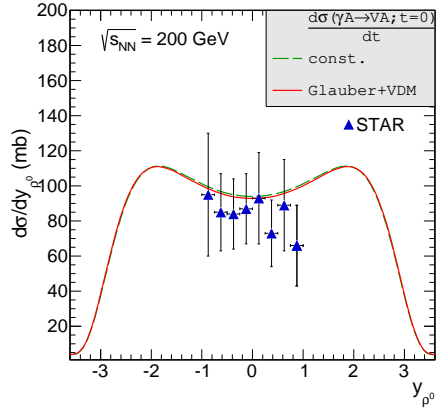}
\end{center}
   \caption{
\small (Color online) Distribution in $\rho^0$ rapidity for single
$\rho^0$. The STAR experimental data are from Ref. \cite{STAR2008_rho0}.
The dashed line is for $d \sigma/dt$ = const and the solid line
is for $d \sigma/dt$ calculated in the Glauber-VDM approach.
}
 \label{fig:dsigma_V_dy}
\end{figure}

Before we go to double $\rho^0$ production, we briefly
show the results for single $\rho^0$ production.
In Fig. \ref{fig:dsigma_V_dy} we present distributions in rapidity.
We obtain similar results as in other calculations in the literature
\cite{KN1999,FSZ2003,GM2005}.
We show rapidity distribution with 
$\frac{d \sigma(\gamma A \to \rho^0 A;t=0)}{dt}$ = 420 mb/GeV$^2$
and result obtained with the Glauber-VDM approach of Ref. \cite{KN1999}
where the elementary cross section was parametrized in a Regge form
with constant (energy-independent) slope.
However, the change turned out to be rather small as it 
is shown in Fig.\ref{fig:dsigma_V_dy}.

It is sufficient for our double-scattering studies that our model 
describes single $\rho^0$ production in the measured region 
of midrapidities. We leave potential
uncertainties related to larger rapidities for future studies.
Our total cross section equals 596 mb.  
The results of models presented in \cite{FSZ2003} and 
\cite{GM2005} exceed the STAR experimental data \cite{STAR2008_rho0}.

%-----------------------------------------------------------------------------
%-----------------------------------------------------------------------------

%-------------------------------------------------------------------------
\subsection{Double scattering mechanism of two $\rho^0$
production}
 %------------------------------------------------------------------------

\begin{figure}[!h]
\begin{center}
\includegraphics[width=5cm]{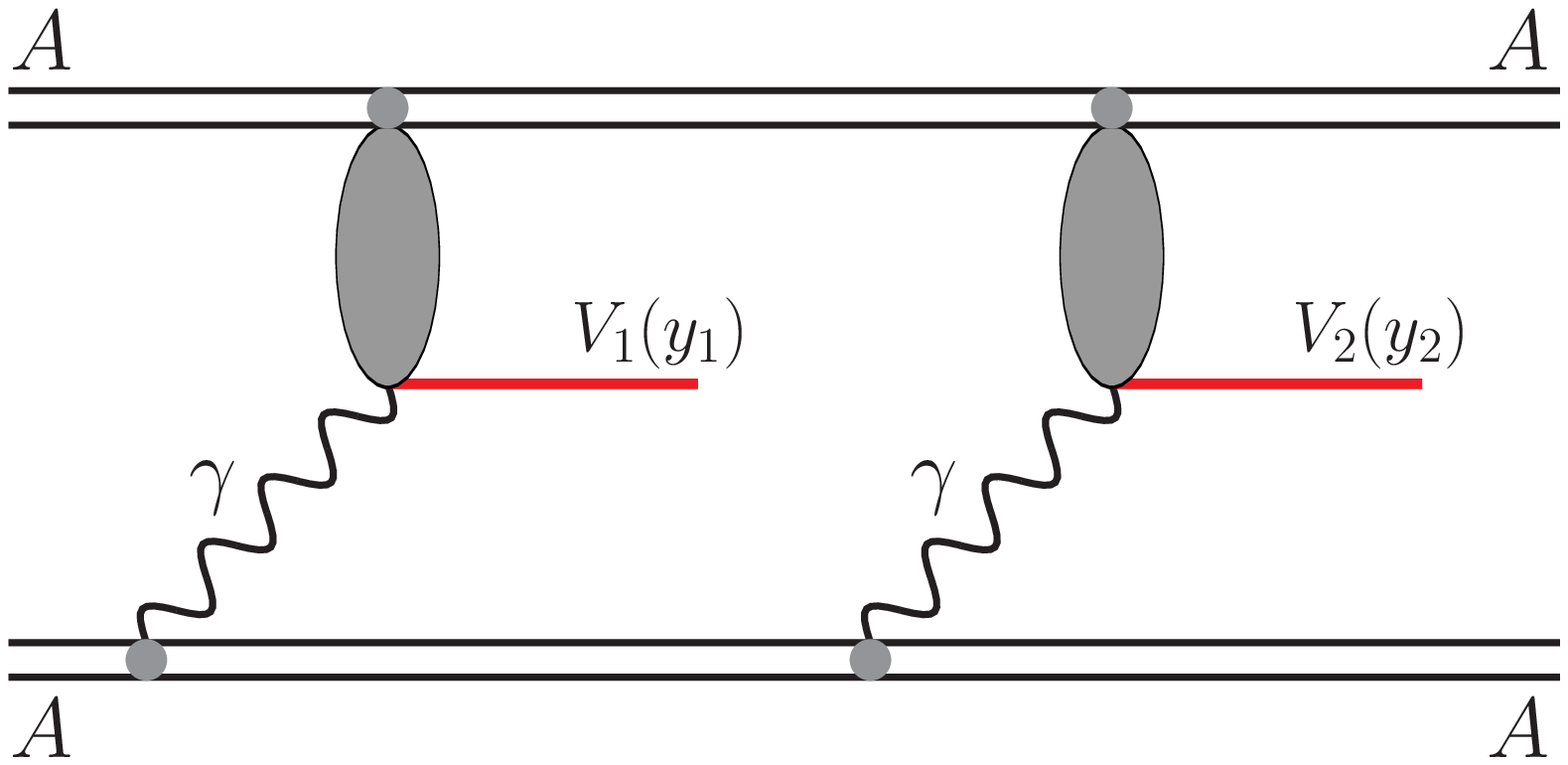}
\includegraphics[width=5cm]{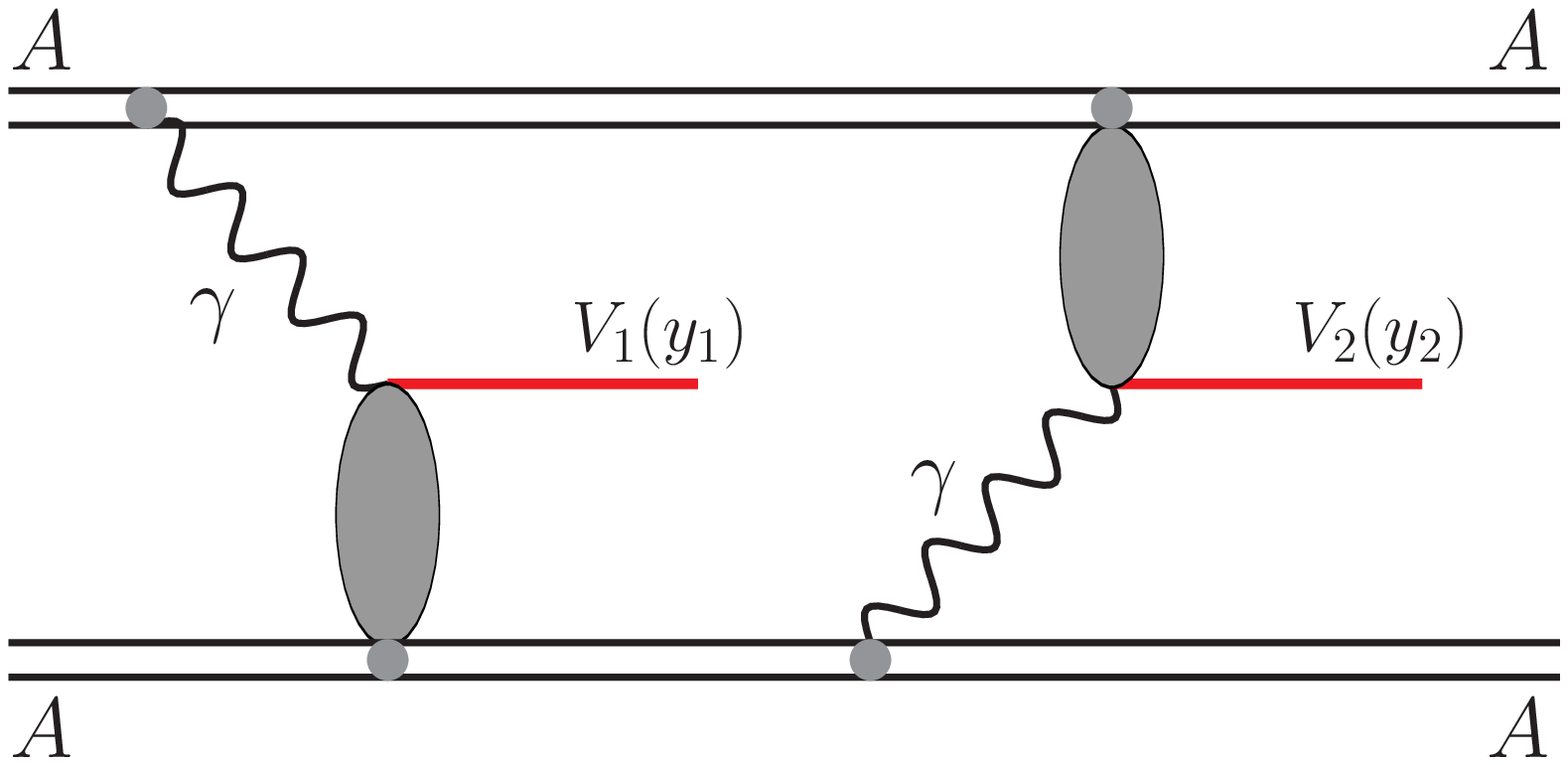}
\\
\includegraphics[width=5cm]{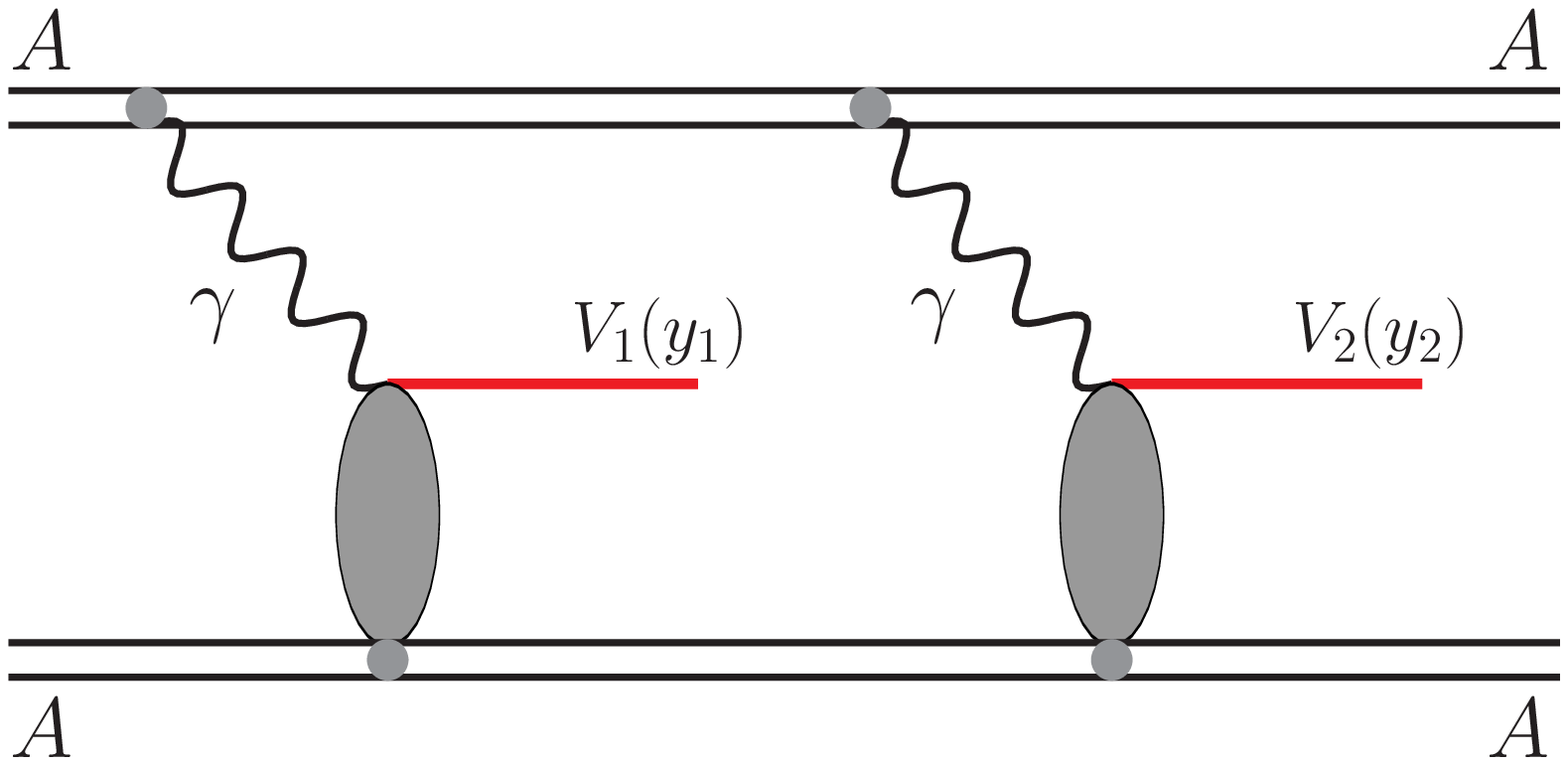}
\includegraphics[width=5cm]{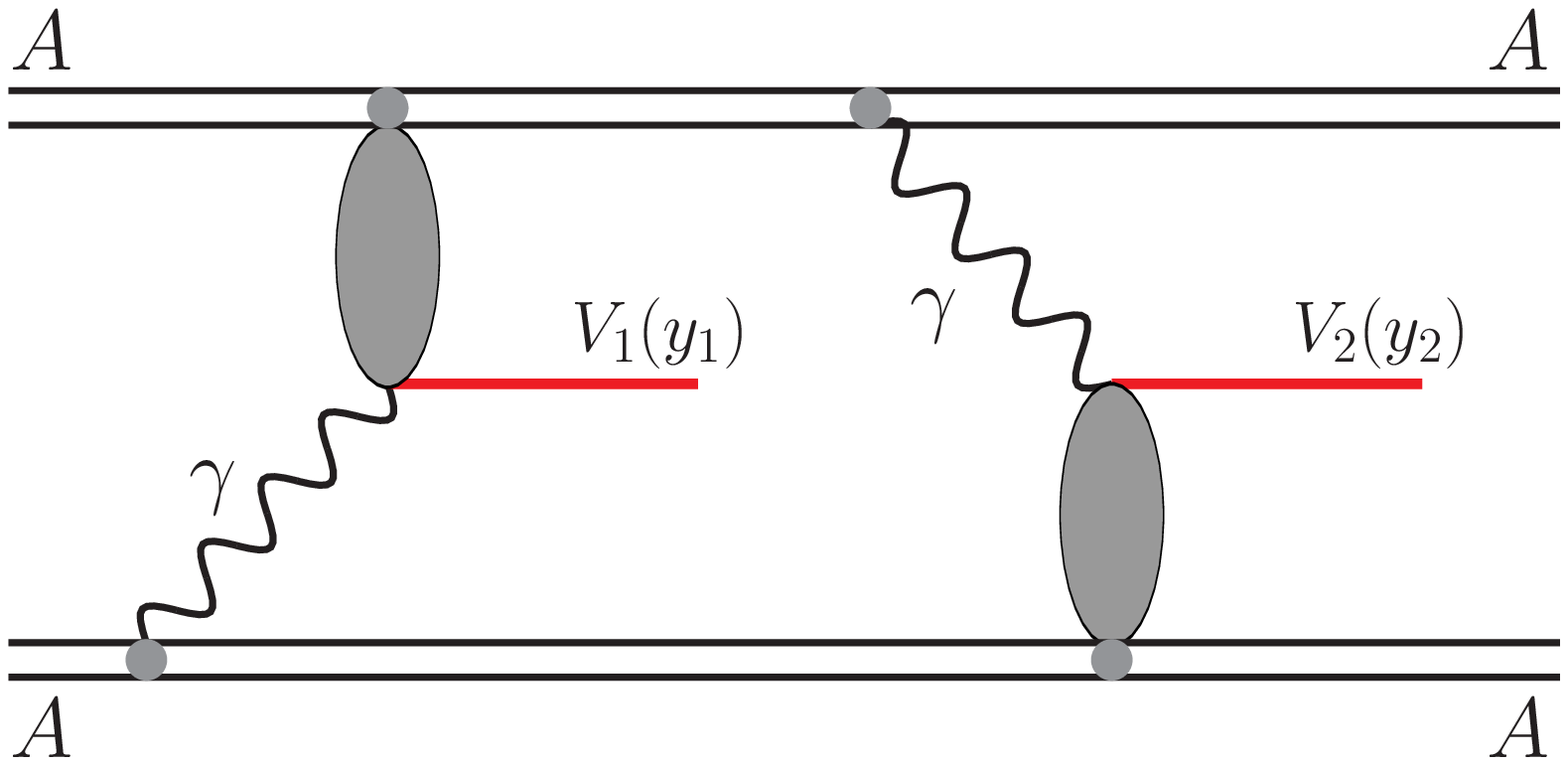}
\end{center}
\caption{
\small (Color online) The double-scattering mechanisms of two vector meson
production.
The blobs denote multiple scattering of quark-antiquark dipoles or
hadronic meson-like photon in the nucleus.
}
\label{fig:double_scattering}
\end{figure}

The generic diagrams of double-scattering production via photon-pomeron
or pomeron-photon exchange \footnote{For brevity and by analogy
to nucleon-nucleon collisions we use the term ``pomeron
  exchange'' which in fact means
complicated high-energy multiple diffractive rescattering of 
quark-antiquark pairs or virtual vector mesons.} 
mechanism are shown in Fig. \ref{fig:double_scattering}.

%-----------------------------------------------------------------------------
%-----------------------------------------------------------------------------

The double scattering process was discussed only in Ref.~\cite{KN1999},
where a probabilistic formula for double and multiple vector meson
production was given. Then the cross section for double
scattering can be written as:
\begin{equation}
\sigma_{A A \to A A V_1 V_2}(\sqrt{s_{NN}}) = 
C \int S_{el}^2(b) P_{V_1}(b,\sqrt{s_{NN}}) P_{V_2}(b,\sqrt{s_{NN}}) d^2 b \; .
\label{sigma_AA_AAV}
\end{equation}
In the equation above $b$ is the transverse distance 
between nuclei. We have included natural limitations in the impact
parameter
\begin{equation}
S_{el}^2(b) = \exp \left( -\sigma_{NN}^{tot} T_{A_1 A_2}(b) \right)
\approx \theta\left( b - (R_1+R_2) \right)  \;  .
\end{equation}
This may be interpreted as a survival probability for nuclei not to
break up.
The probability density of single vector meson production is
\begin{equation}
P_V(b,\sqrt{s_{NN}}) = \frac{d \sigma_{AA \to AAV}(b;\sqrt{s_{NN}})}
{2 \pi b db} \; .
\label{probability}
\end{equation}
The constant $C$ is in the most general case 1 or $\frac{1}{2}$ for 
identical vector mesons $V_1 = V_2$.
We have explicitly indicated the dependence of the probabilities on 
nucleon-nucleon energy.
The probability densities $P_V$ increase with increasing cm energy.

Here the photon flux factor is calculated as:
\begin{equation}
\frac{d^3N}{d^2b d\omega} 
= \frac{Z^2\alpha_{em}X^2}{\pi^2\omega b^2} K_1^2(X)  \; ,
\end{equation}
where 
$X = \frac{b \omega}{\gamma}$.
 
The simple formula (\ref{sigma_AA_AAV}) can be generalized to calculate 
two-dimensional distributions in rapidities of both vector mesons
\begin{eqnarray}
\frac{d \sigma_{AA \to AA V_1 V_2}}{d y_1 d y_2} = C \int 
 && \left( \frac{d P_1^{\gamma \Pom}(b,y_1;\sqrt{s_{NN}})}{d y_1}
               + \frac{d P_1^{\Pom \gamma}(b,y_1;\sqrt{s_{NN}})}{d y_1}
               \right)
\nonumber \\
\times 
 && \left( \frac{d P_2^{\gamma \Pom}(b,y_2;\sqrt{s_{NN}})}{d y_2}
                + \frac{d P_2^{\Pom \gamma}(b,y_2;\sqrt{s_{NN}})}{d y_2}
               \right)   \;  d^2 b  \; . 
\label{dsigma_dy1dy2}    
\end{eqnarray}
$P_1$ and $P_2$ are probability densities for producing
one vector meson $V_1$ at rapidity $y_1$ and the second vector meson $V_2$
at rapidity $y_2$ for the impact parameter $b$. 
Then the differential probability density reads: 
\begin{equation}
\frac{d P_V(b,\sqrt{s_{NN}})}{dy} =
 \frac{d \sigma_{AA \to AAV}(b;\sqrt{s_{NN}})}
{2 \pi b db dy} \; .
\label{generalized_probability}
\end{equation}
The produced vector mesons in each step are produced in very 
broad range of (pseudo)rapidity \cite{KN1999,GM2006} and extremely
small transverse momenta.

%--------------------
\section{Results}
%--------------------

%-------------------------------------------------------------
\subsection{$Pb Pb \to Pb Pb \pi \pi$}
\begin{figure}[!h]             
\begin{center}
\includegraphics[width=6cm]{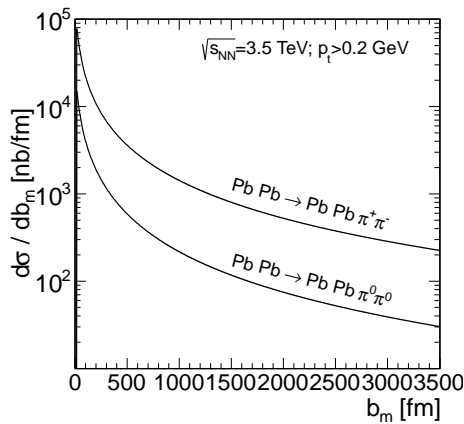}
\end{center}
\caption{\label{fig:dsig_dbm}
\small (Color online) The impact parameter distribution for 
the $Pb Pb \to Pb Pb \pi^+ \pi^-$ and $Pb Pb \to Pb Pb \pi^0 \pi^0$
reactions at $\sqrt{s_{NN}} = $ 3.5 TeV.}
\end{figure}

%-------------------------------------------------------------
%
We start presentation of our results in Fig.\ref{fig:dsig_dbm}
by showing interesting distribution in impact parameter between
the two lead nuclei. The distribution is different from zero 
starting from the distance
of $b = 2 R_A \approx$ 14 fm and extends till ''infinity''. 
This means that a big part of the cross section comes from the situations 
when the two $^{208}Pb$ nuclei fly far one from the other. This
demonstrates the ultraperipheral character of the discussed 
process of exclusive dipion production via photon-photon fusion.

%-----------------------------------------------------------------------------
%-----------------------------------------------------------------------------

%------------------------------------------------------------------------
\begin{figure}[!h]             
\begin{center}
\includegraphics[width=6cm]{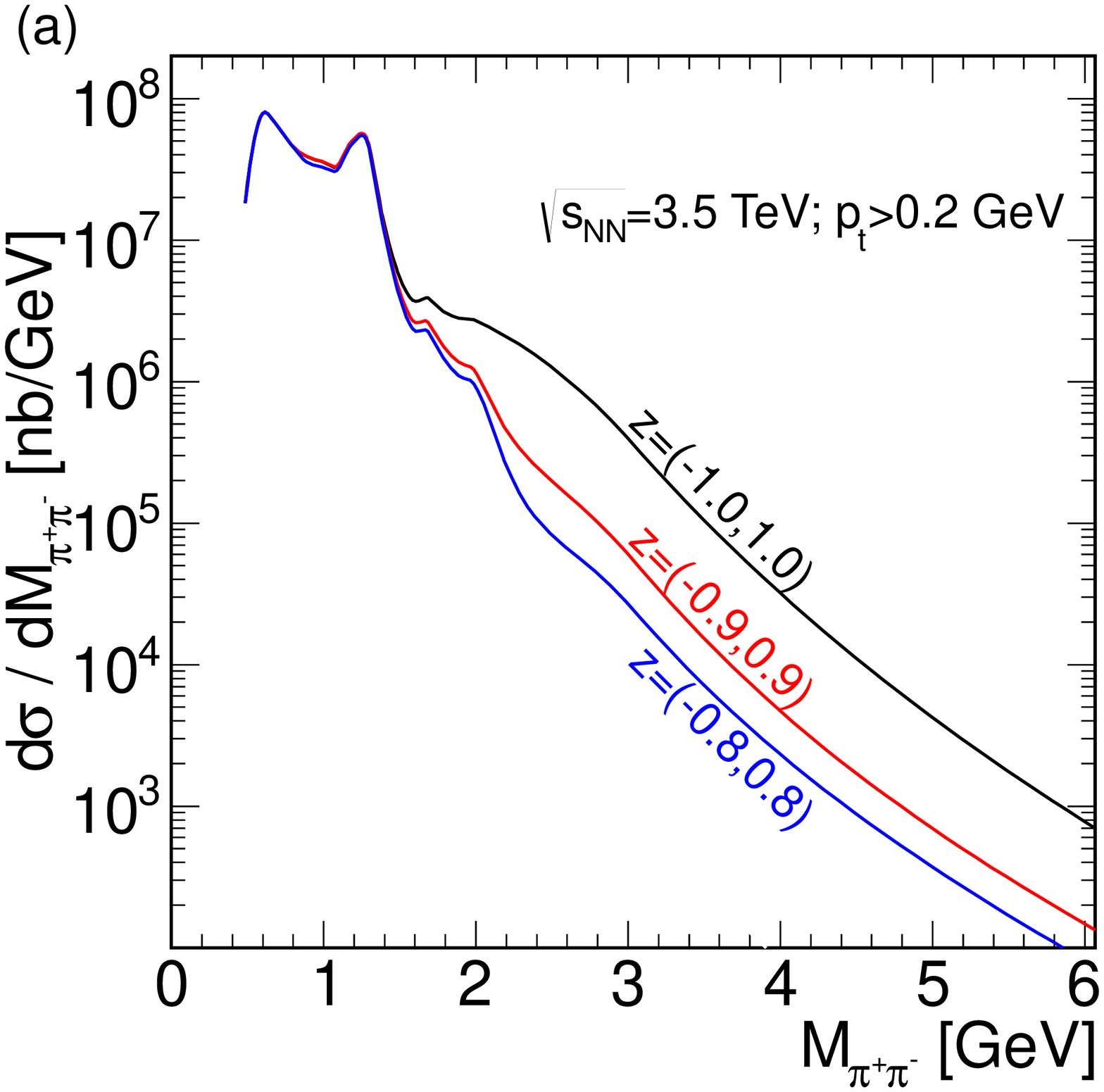}
\includegraphics[width=6cm]{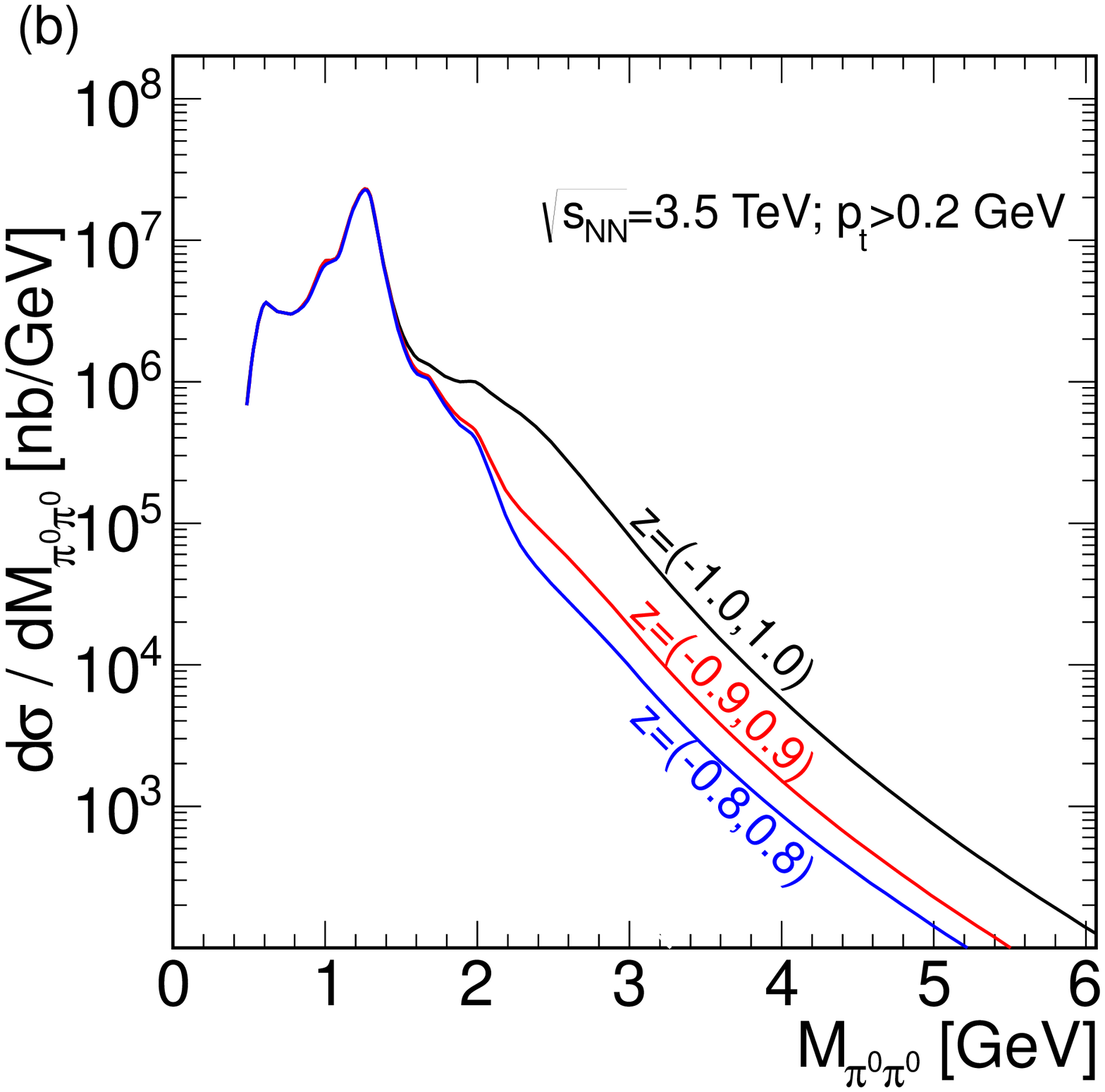}
\end{center}
   \caption{
\small  (Color online) Dipion invariant mass for 
$\pi^+ \pi^-$ (left panel) and 
$\pi^0 \pi^0$ (right panel) at the LHC energy $\sqrt{s_{NN}} =$ 3.5 TeV
and $p_{t,\pi} >$ 0.2 GeV. 
We show also results with extra cuts on $z$.}
\label{fig:dsig_dW}
\end{figure}
%------------------------------------------------------------------------
%

The distribution in invariant dipion mass is shown in Fig.\ref{fig:dsig_dW}.
The distribution for the full phase space (upper solid line) is shown
for the $PbPb \to PbPb\pi^+\pi^-$ (left panel) and 
for the $PbPb \to PbPb\pi^0\pi^0$ (right panel) reactions
for the full phase space
(upper lines), for $|\cos\theta|<$ 0.9 (middle lines) and 
for $|\cos\theta|<$ 0.8 (lowest line).
At lower dipion invariant masses the result does not depend on 
the angular cuts.

%------------------------------------------------------------------------
\begin{figure}[!h]             
\begin{center}
\includegraphics[width=6cm]{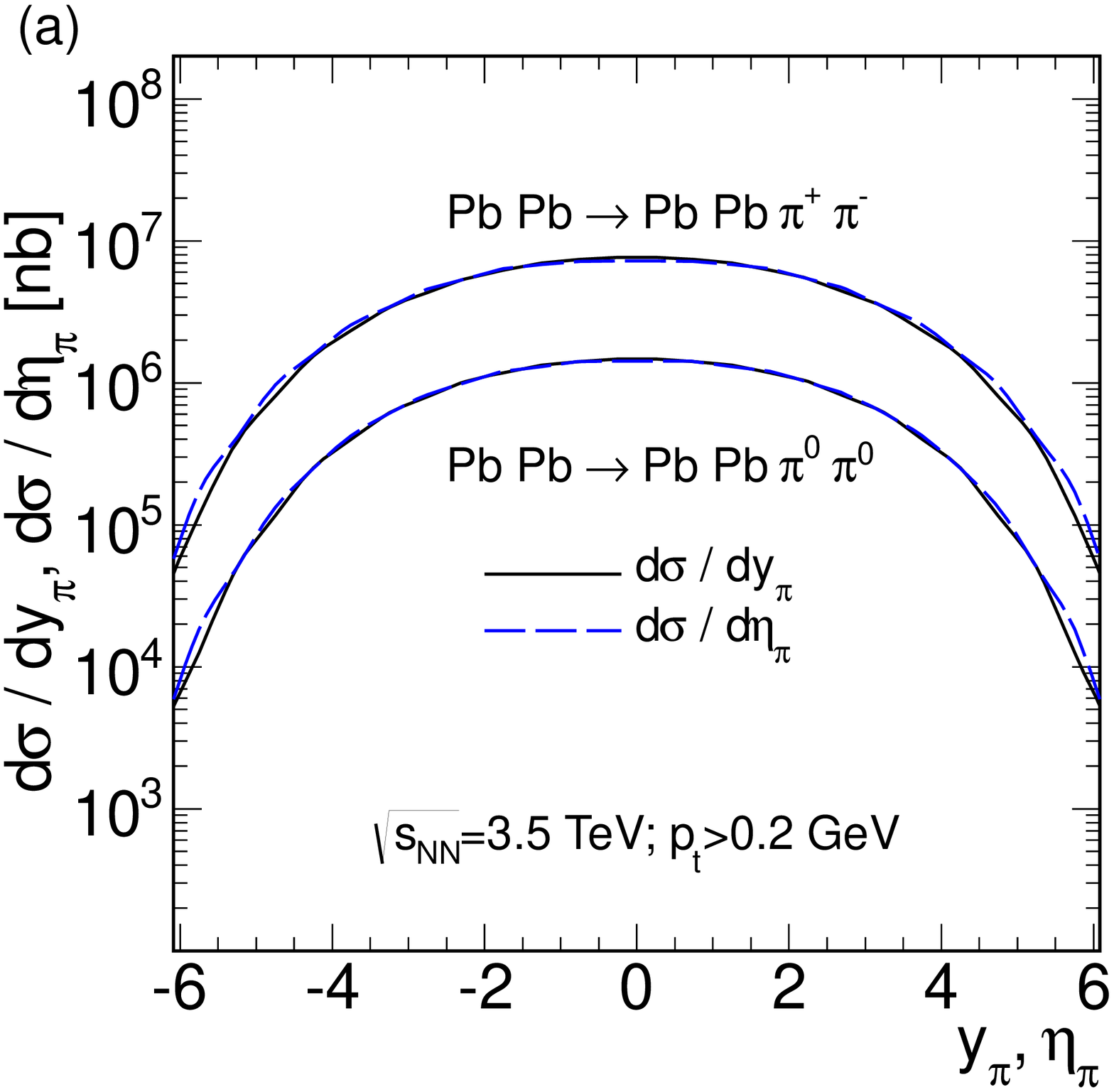}
\includegraphics[width=6cm]{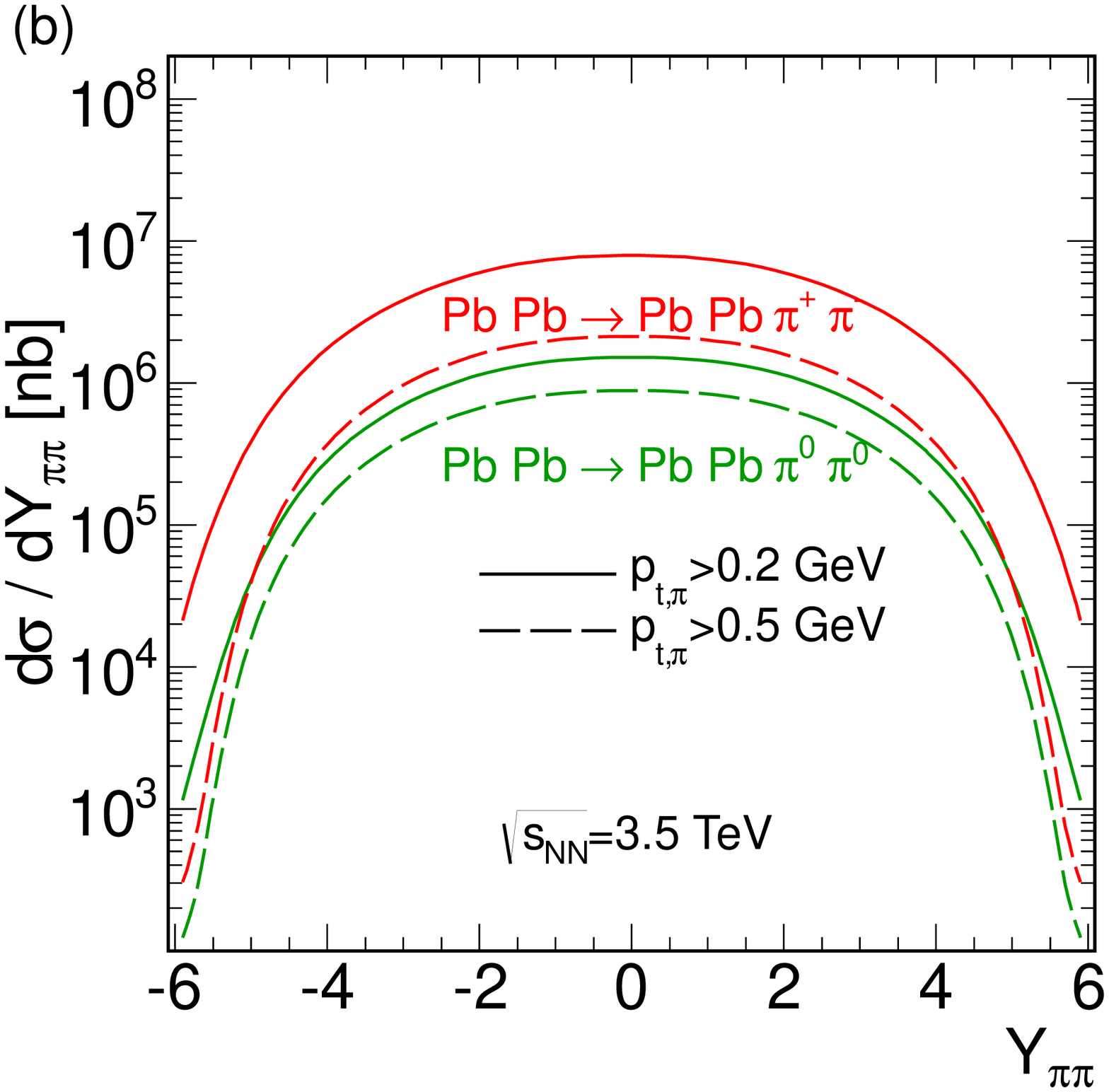}
\end{center}
   \caption{\label{fig:dsig_dy_deta}
   \small  (Color online) 
$\frac{d\sigma}{dy_{\pi}}$ and $\frac{d\sigma}{d\eta_{\pi}}$ (left panel)
and $\frac{d\sigma}{dY_{\pi\pi}}$ (right panel) for the LHC energy 
$\sqrt{s_{NN}} =$ 3.5 TeV.}
\end{figure}
%------------------------------------------------------------------------

Let us start now presentation of theoretical differential distributions
that can be measured.
In the left panel of Fig.\ref{fig:dsig_dy_deta} we show distributions 
in rapidity of individual pions (solid line) and 
distribution in pseudorapidity of the same pion (dashed line). 
Right panel illustrates nuclear cross section as the function of 
the pion pair rapidities.
Here we have imposed extra cuts on pion transverse momenta:
$p_{t,\pi}>$ 0.2 GeV (solid lines) and $p_{t,\pi}>$ 0.5 GeV (dashed lines).
In addition, we compare the nuclear cross section 
for $\pi^+\pi^-$ (upper curves) and for $\pi^0\pi^0$ (lower curves) production.

%========================================================================
\begin{figure}[!h]             
\begin{center}
\includegraphics[width=7cm]{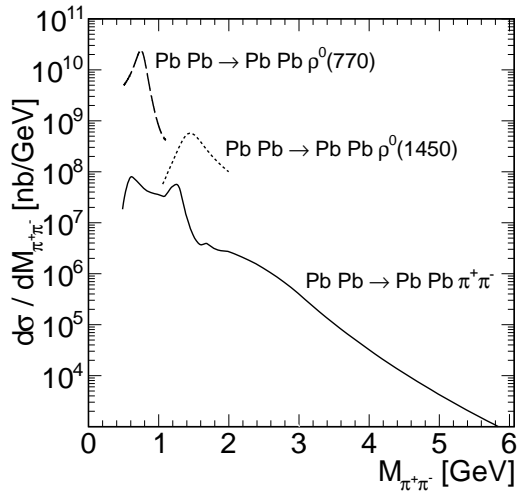}
\end{center}
   \caption{\label{fig:dsig_dW_rho}
   \small  Invariant mass distribution of $\pi^+\pi^-$ from the decay of 
   $\rho^0(770)$ and
   $\rho^0(1450)$ photo-production (resonance contributions represented by 
   the dashed and dotted lines) and our $\gamma\gamma$ fusion (solid line) 
   in ultraperipheral Pb-Pb collisions at $\sqrt{s_{NN}}=$ 3.5 TeV.}
\end{figure}
%========================================================================

In Fig.\ref{fig:dsig_dW_rho} the resonance contribution for the
$\rho^0(1450)$ is 5/0.27 times smaller than for $\rho^0(770)$ 
as suggested by a calculation in the literature and the above upper limit 
for the branching fraction into pions.
It is clear that this is an upper estimate
for the $\rho^0(1450)$ contribution. 

%--------------------------------------------------------------------------
\subsection{Exclusive production of $\rho^0 \rho^0$ pairs}
%--------------------------------------------------------------------------

Now we shall discuss production of $\rho^0 \rho^0$ pairs.

%-----------------------------------------------------------------------------
\begin{figure}[!h]
\begin{center}
\includegraphics[width=6cm]{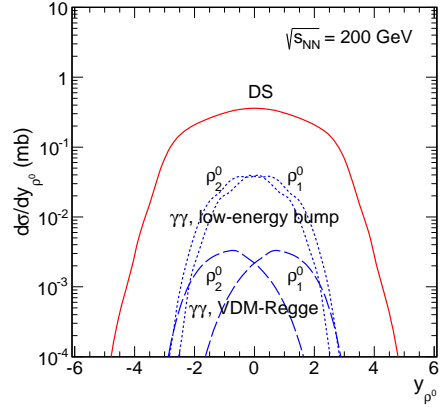}
\end{center}
   \caption{
\small (Color online) Rapidity distribution of one of $\rho^0$ mesons
produced in double scattering mechanism. The double-scattering
contribution is shown by the solid (red online) line and the dashed
lines (blue online) represents distributions of 
$\rho^0$ produced in the high-energy VDM-Regge photon-photon fusion.}
 \label{fig:dsig_dyrho}
\end{figure}
%-----------------------------------------------------------------------------

Distributions in $\rho^0$ meson rapidity is shown 
in Fig.~\ref{fig:dsig_dyrho}. One observes a dominance
of the double scattering component over the photon-photon component. 
At the LHC the proportions will be slightly modified.
For the photon-photon mechanism we show separate contributions 
for the forward and backward $\rho^0$ mesons.

%-----------------------------------------------------------------------------
\begin{figure}[!h]
\begin{center}
\includegraphics[width=6cm]{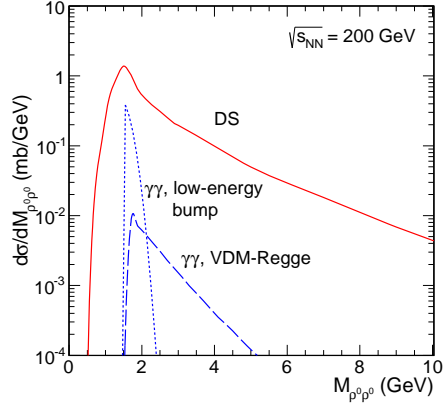}
\end{center}
   \caption{
\small (Color online) Invariant mass distribution of $\rho^0 \rho^0$ for double
scattering (solid line), high-energy VDM-Regge photon-photon (dashed line)
and low-energy bump (dotted line) contributions
for full phase space.}
\label{fig:dsig_dMVV}
\end{figure}
%------------------------------------------------------------------------------

The corresponding distribution in the $\rho^0 \rho^0$ invariant mass is 
shown in Fig.~\ref{fig:dsig_dMVV}. We show both low-energy and
high-energy photon-photon contributions. The low energy component is a
purely mathematical fit from Ref.\cite{KS_rho}. 
This may be a bit an artifact of a simple functional form used. 
The issue is a bit difficult as the peak appears close to the
threshold.
This could be also some close to threshold mechanism.
Our purely matematical representation of the uknown effect is 
therefore oversimplified.
In general, larger invariant masses
are generated via the double scattering mechanism than in two-photon 
processes. 
The reader is asked to compare the present plot with analogous plot in 
Ref.~\cite{KS_rho}.

%-----------------------------------------------------------------------------
\begin{figure}[!h]
\begin{center}
\includegraphics[width=6cm]{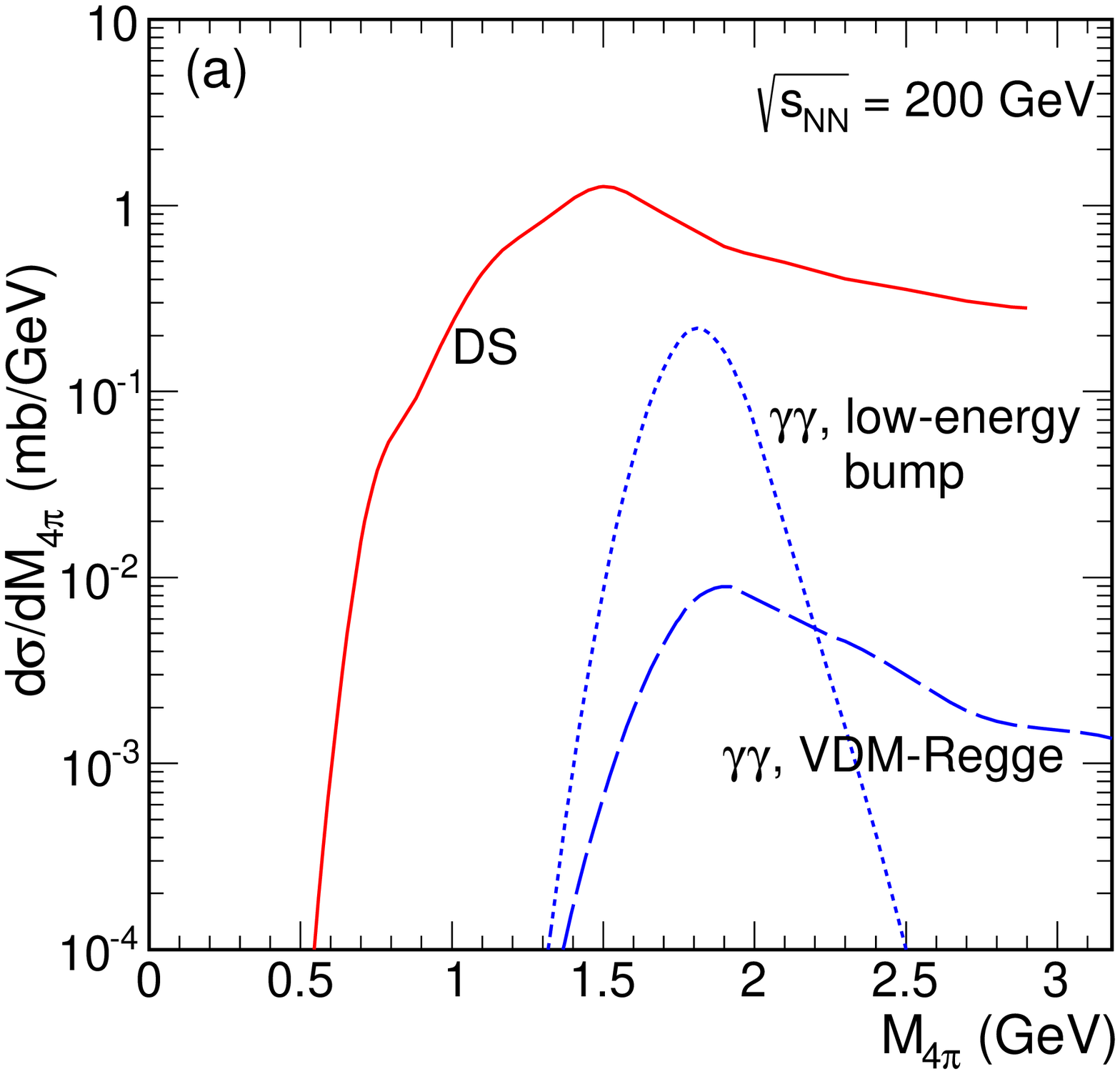}
\includegraphics[width=6cm]{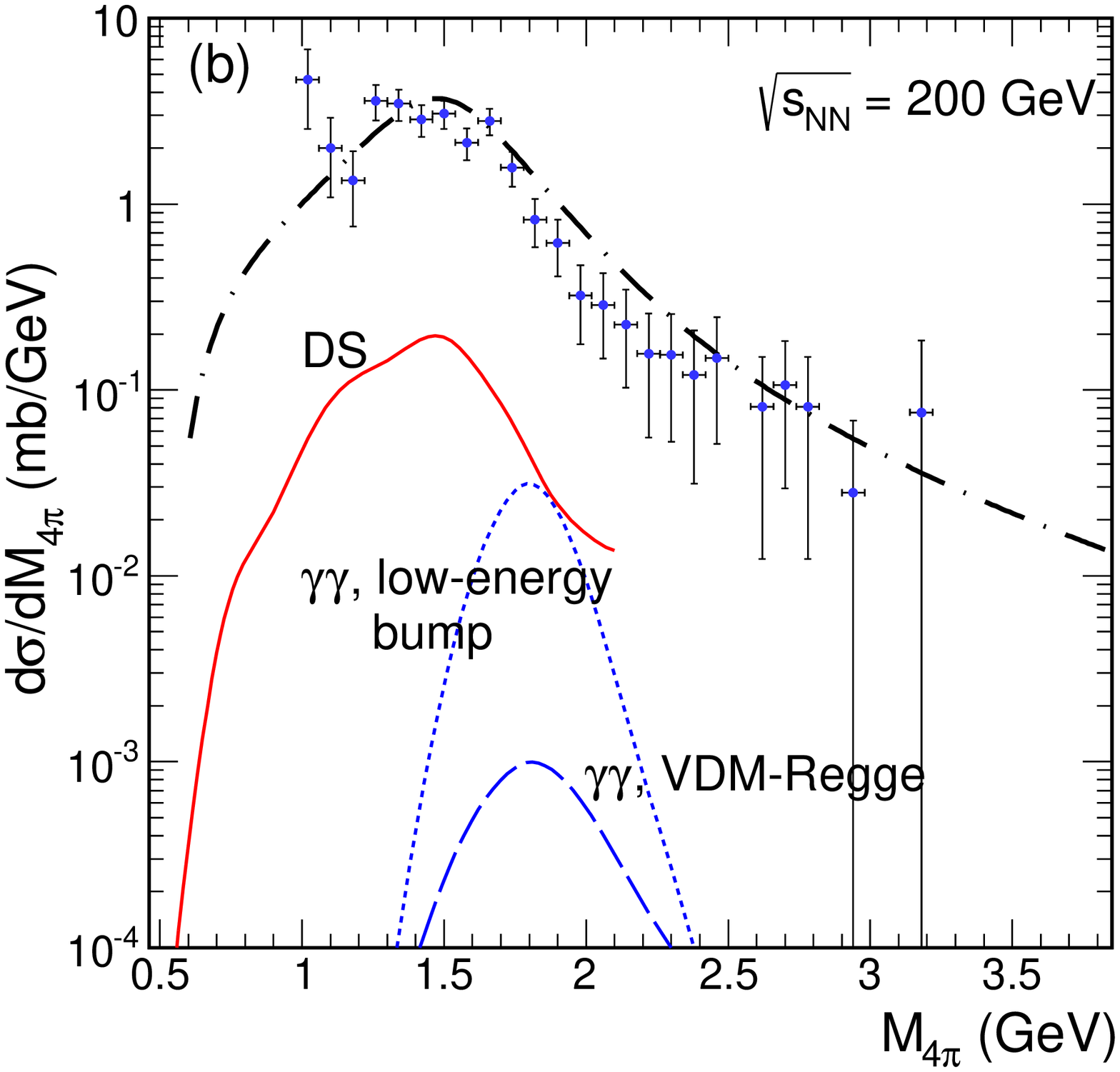}
\end{center}
   \caption{
\small (Color online) Four-pion invariant mass distribution for double
scattering mechanism (solid line), high-energy VDM-Regge photon-photon 
(dashed line) and low-energy bump (dotted line) mechanisms 
for full phase space (left panel) and for the limited acceptance 
STAR experiment (right panel). 
The STAR experimental data \cite{STAR2010_4pi} have been corrected
by acceptance function \cite{BGpc}. The dash-dotted line represents
a fit of the STAR collaboration.}
 \label{fig:dsig_dM4pi}
\end{figure}
%------------------------------------------------------------------------------

In experiments, charged pions are measured rather than $\rho^0$ mesons.
Therefore, we now proceed to a presentation of some observables related 
to charged pions. In Fig.~\ref{fig:dsig_dM4pi} we present four-pion
invariant mass distribution. The distribution for the
whole phase space extends to large invariant masses, while the
distribution in the limited range of (pseudo)rapidity as defined by the STAR
detector give a shape similar to the measured distribution (see
dash-dotted line in the right panel of Fig. \ref{fig:dsig_dM4pi}).
However, the double-scattering contribution accounts only for 20 \% of 
the cross section measured by the STAR collaboration \cite{STAR2010_4pi}. 
Probably, the production of the $\rho^0(1700)$ resonance and its 
subsequent decay into the four-pion final state (see e.g., \cite{PDG}) 
is the dominant effect for the limited STAR acceptance. Both, the production 
mechanism of $\rho^0(1700)$ and its decay into four charged pions are 
not yet fully understood. There was only one attempt to calculate
the production cross section in the Glauber-Gribov GVDM approach
\cite{FSZ2003b}. Furthermore there is another broad
$\rho(1450)$ resonance \cite{PDG} which also decays into four charged pions. 
We leave the modeling of the production and decay processes for 
a dedicated study.

%-----------------------------------------------------------------------------
\begin{figure}[!h]
\begin{center}
\includegraphics[width=6cm]{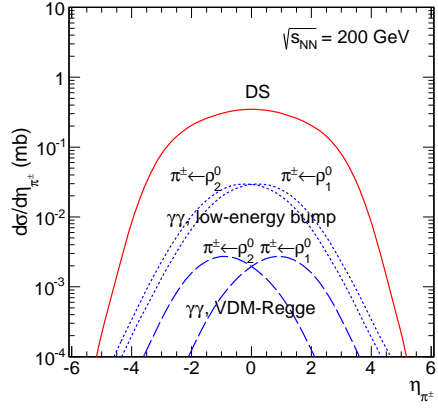}
\end{center}
   \caption{
\small (Color online) Pseudorapidity distribution of charged pions
for double scattering (solid line), high-energy VDM-Regge photon-photon 
(dashed line) and low-energy bump (dotted line) mechanisms 
for full phase space.
}
 \label{fig:dsig_detapi}
\end{figure}
%------------------------------------------------------------------------------

In Fig.~\ref{fig:dsig_detapi} we show distributions in pseudorapidity of 
the charged pions. The distributions extend over a broad range
of pseudorapidity. Both STAR collaboration at RHIC and the ALICE 
collaboration at LHC can observe only a small fraction of pions 
due to the rather limited pseudorapidity coverage. 
While the CMS pseudorapidity coverage is wider, 
it is not clear if the CMS collaboration has a relevant trigger 
to measure the exclusive nuclear processes.

%-----------------------------------------------------------------------------
\begin{figure}[!h]
\begin{center}
\includegraphics[width=6cm]{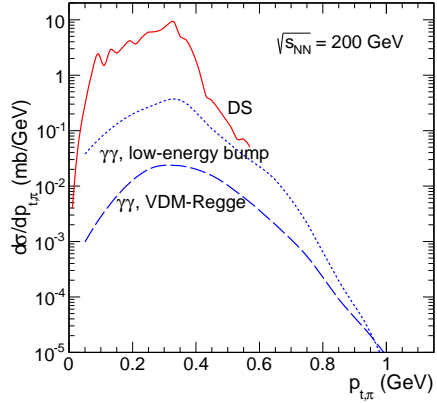}
\end{center}
   \caption{
\small (Color online) Transverse momentum distribution of charged pions
for double scattering (solid line) low-energy bump (dotted line)
and high-energy VDM-Regge photon-photon (dashed line)
mechanisms for full phase space.
}
 \label{fig:dsig_dptpi}
\end{figure}
%------------------------------------------------------------------------------

For completeness in Fig.~\ref{fig:dsig_dptpi} we show distributions
in pion transverse momenta. 
%The distributions at RHIC and LHC have quite similar shapes. 
Since the $\rho^0$ mesons produced in the double-scattering
photon-induced mechanisms have very small transverse momenta, 
the transverse momenta of pions are limited to $\sim m_{\rho^0}/2$. 
The distribution is relatively smooth,
because here we have taken into account a smearing of $\rho^0$ meson masses. 
The sharp upper limit is an artifact of our maximal value of $\rho^0$
meson mass $m_{\rho}^{max}$ = 1.2 GeV. We have imposed this upper limit
because the spectral shape of ``$\rho^0$ meson'' above 
$m_{\rho} >$ 1.2 GeV is not well known.
At larger $p_{t,\pi}$, the contribution from 
the decay of $\rho^0$ meson produced in photon-photon 
fusion can be larger as that of double scattering mechanism, as 
the transverse momentum of $\rho^0$ mesons are not strictly limited 
to small values. However, the cross section for such cases is expected 
to be very small. Both STAR ($p_t >$ 0.1 GeV) and ALICE ($p_t >$ 0.1 GeV) 
experiments have a fairly good coverage in pion transverse momenta
and could measure such distributions.

%--------------------------
\section{Conclusions}
%--------------------------

In Ref.\cite{KS2013} we have calculated total cross sections and 
angular distribution as a function of $\gamma \gamma$ energy for both 
$\gamma \gamma \to \pi^+ \pi^-$ and $\gamma \gamma \to \pi^0 \pi^0$
processes. These energy-dependent 
cross sections have been used in Equivalent Photon 
Approximation in the impact parameter space to calculate 
corresponding production rate in ultraperipheral
ultrarelativistic heavy ion reactions. In this calculation we have taken
into account realistic charge distributions in colliding nuclei.

We have calculated both total cross sections at LHC energy, 
and distributions in rapidity and transverse momentum of pions
and dipion invariant mass. The calculation of distributions of 
individual pions in the b-space EPA is more complicated. 
The distributions in dipion invariant mass have been compared
with the contribution of exclusive $\rho^0 \to \pi^+ \pi^-$ production
in photon-pomeron or pomeron-photon mechanism.
Close to the $\rho^0$ resonance the 
$\gamma \gamma \to \pi^+ \pi^-$ mechanism yields only a small
contribution. We hope the $\gamma \gamma$ contribution could be 
identified or even measured outside of the $\rho^0$ resonance region. 
A detailed comparison with the
absolutely normalized experimental data of the ALICE collaboration should 
allow a test of our predictions.
We have discussed two-$\rho^0$  as well as four-pion
production in exclusive ultraperipheral heavy ion collisions,
concentrating on the double scattering mechanism.

Differential distributions for the two $\rho^0$ mesons and
for four pions have been presented. The results for total cross section 
and differential distributions for the double scattering mechanism 
have been compared with the results for two-photon fusion discussed 
already in the literature.
We have found that at $\sqrt{s_{NN}}$ = 200 GeV 
the contribution of double scattering is almost two orders of magnitude 
larger than that for the photon-photon mechanism.

The produced $\rho^0$ mesons decay, with large probability, 
into charged pions.
In the consequence this leads to large contribution to exclusive 
production of the $\pi^+ \pi^- \pi^+ \pi^-$ final state.
We have made a comparison of four pion production via $\rho^0 \rho^0$
production (double scattering and photon-photon fusion) 
with experimental data measured by the STAR collaboration for
gold-gold scattering.  
The theoretical predictions have fairly similar shape in four-pion invariant
mass distribution as measured by the STAR collaboration 
but exhaust only a quarter of the measured cross section. 
The missing contribution is probably
due to  the exclusive production of $\rho^0(1700)$ resonance 
and its decay into four charged pions. 
We expect that in the total phase space the contribution of double scattering
is similar to that for the $\rho^0(1700)$ resonant production.

A separation of double scattering, photon-photon and $\rho'$ mechanisms
seems very important. In general, transverse momentum of each
of the produced $\rho^0$'s in double scattering mechanism is very small,
smaller than in the other mechanisms. As a consequence the pions from 
the decay of $\rho^0$'s from the double-scattering mechanism are 
produced back-to-back in azimuthal angle.
This could be used to enhance the signal of double scattering mechanism.
At large pseudorapidity separations between two $\rho^0$'s and/or 
large $\pi^+ \pi^+$ ($\pi^- \pi^-$) pseudorapidity
separations the double scattering contribution should dominate
over other contributions. The identification of the region 
seems difficult at RHIC but could be better at the LHC.

At present the four charged pion final state is being analyzed 
by the ALICE collaboration. We plan a separate careful analysis for 
the ALICE and other LHC experiments. It would be valuable if the different
mechanisms discussed in the present paper could be separated
experimentally in the future. This requires, however, 
rather complicated correlation studies for four charged pions. 

Similar double scattering mechanisms could be studied for different
vector meson production, e.g. for $\rho^0 J/\Psi$ production. 
Recently we have sudied production of $J/\Psi J/\Psi$ pairs via 
two-photon mechanism \cite{BCKSS2013}. A calculation
of the corresponding double-scattering contribution would be in this
case very interesting.

\vspace{1cm}

{\bf Acknowledgments}

This presentation is based on common work with
Mariola K{\l}usek-Gawenda and Wolfgang Sch\"afer.
This work was partially supported by 
N DEC-2011/01/B/ST2/04535.

%uncomment the following lines to place a figure
%\begin{figure}[htb]
%\centerline{%
%\includegraphics[width=12.5cm]{Fig1}}
%\caption{Plot of ...}
%\label{Fig:F2H}
%\end{figure}

\end{document}